\documentclass[manuscript]{aastex}

\newcommand{\myemail}{aya@shibaura-it.ac.jp}

\shorttitle{truncated disk in the VHS}
\shortauthors{Tamura et al.}

\begin{document}


\title{The truncated disk from Suzaku data of GX~$339-4$ in the extreme very high state }
\author{Manami Tamura and Aya Kubota}
\affil{Department of Electronic Information Systems, Shibaura Institute of Technology, \\
307 Fukasaku, Minuma-ku, Saitama, Saitama 337-8570 Japan}
\email{m110090@shibaura-it.ac.jp, \myemail}

\author{Shinya Yamada}
\affil{Institute of Physical and Chemical Research (RIKEN), 2-1 Hirosawa, Wako, Saitama 351-0198, Japan}
\author{Chris Done and  Mari Kolehmainen}
\affil{Department of Physics, University of Durham, 
South Road, Durham, DH1 3LE, UK}
\author{Yoshihiro Ueda}
\affil{Department of Astronomy, Kyoto University, Kitashirakawa-Oiwake-cho, Sakyo-ku, Kyoto, Kyoto 606-8502, Japan}
\and
\author{Shunsuke Torii}
\affil{Department of Physics, The University of Tokyo, 7-3-1 Hongo, Bunkyo-ku, Tokyo 113-0033, Japan}



\begin{abstract}

We report on the geometry of accretion disk and high energy coronae in
the strong Comptonization state (the
very high/steep power law/hard intermediate 
state) based on a Suzaku
observation of the famous Galactic black hole GX~$339-4$. 
These data were taken just before the peak of the 2006--2007 outburst, 
and
the average X-ray luminosity in the 0.7--200~keV band is estimated to be
$2.9\times 10^{38}~{\rm erg~s^{-1}}$ for a distance of 8~kpc.  
We fit
the spectrum with both simple (independent disk and corona)
and sophisticated (energetically coupled disk and corona) models, but all
fits imply that the underlying optically thick disk is truncated
significantly before the innermost stable circular orbit around the
black hole.  We show this directly by a comparison with similarly
broadband data from a disk dominated spectrum at almost the same
luminosity observed by XMM-Newton and RXTE 3 days after the Suzaku
observation.
During the Suzaku observation, the QPO frequency changes from 4.3~Hz
to 5.5~Hz, while the spectrum softens. The energetically coupled
model gives a corresponding $5\pm 8$\% decrease in derived inner radius of the 
disk. While this is not significant, it is consistent with the predicted change in 
QPO frequency from Lense-Thirring precession of the hot flow interior to the disk
and/or a deformation mode of this flow,
as a higher QPO frequency implies a smaller
size scale for the corona.
This is consistent with the truncated disk extending further
inwards towards the black hole.

\end{abstract}



\section{Introduction}

The accretion flow in Black Hole Binaries (BHB) is generically unstable, so most sources
are transient  (e.g. \citealt{lasota01}), 
showing dramatic outbursts with a large change in mass accretion rate
and a correspondingly large change in X-ray properties \citep{mr06}.
Generally these start in the low/hard state (LHS) where the disk is very dim, and the spectrum
is dominated by a hard Comptonized spectrum up to 100--200~keV. This brightens rapidly
while remaining hard, then there is an abrupt change, where the disk spectrum 
strengthens rapidly and the Compton tail softens to 
$\Gamma\ge 2.0$.
The tail can then become quite weak,
so that the spectrum is dominated by the disk emission with temperature $\le 1$~keV
of which 
the Wien tail 
extends into  the RXTE bandpass above 3~keV. 
This state is called the high/soft state (HSS).
The source then remains in the HSS as its 
luminosity declines, then makes a transition back to the LHS. This typically happens at a lower
luminosity than the LHS to HSS transition, a hysteresis effect which gives a characteristic $q$
shape on a hardness-intensity diagram of the outburst (see e.g., the compilation of 
\citealt{dunn10}). 

In the HSS,  the disk emission can be fit by the {\sc diskbb} 
model which approximates the standard disk~\citep{ss73} by  
describing the local disk temperature as 
$T(r) = T_{\rm in}\cdot (r/r_{\rm in})^{-3/4}$~\citep{mitsuda84} with
the maximum observed disk temperature $T_{\rm in}$ and an apparent disk inner radius $r_{\rm in}$.
In this modeling, the disk
bolometric luminosity $L_{\rm disk}$ can be related to these two spectral parameters
as $L_{\rm disk} =4\pi r_{\rm in}^2\sigma T_{\rm in} ^4$.
Compelling evidence for the standard disk formalism is given by the observation that the value of $r_{\rm in}$
remains remarkably constant in the HSS as $L_{\rm disk}$ changes significantly.
Thus, after several corrections including the stress-free inner boundary condition (e.g., \citealt{kubota98,gier99}), 
color temperature correction \citep{shimura95}, and relativistic corrections \citep{cunningham75, zcc97},
$r_{\rm in}$ is generally believed to be consistent with the  
innermost stable circular orbit (ISCO) around central black holes.

By contrast, in the LHS the disk emission is very weak and the luminosity is
instead dominated by the hard ($\Gamma<2$) power law tail. The radius of the disk is
hard to measure directly in this state due to model uncertainties as well as the weakness
of the disk emission (see e.g \citealt{dgk07}). Models of this 
state generally assume that the inner
disk progressively recedes at low luminosities,  being replaced by the alternative
hot inner flow solutions of the accretion flow equations
(e.g., \citealt{esin97}). These models then predict that as the source makes 
a transition from the HSS towards the LHS, the disk should
start to pull back from the ISCO. This cannot be observed directly in RXTE data 
due to the 3~keV lower limit to the bandpass. However, there is one
source monitored by SWIFT during a transition, and here the
radius of the thermal disk component clearly increases as predicted as the 
the source declines from a HSS into the intermediate state \citep{gdp08}, although the
LHS radius is much more model dependent \citep{rykoff07,gdp08}.

The intermediate states from LHS to HSS on the rise often appear quite similar to the spectra
seen as the source makes a transition from HSS to LHS on the decline, but at higher
luminosity due to the hysteresis noted above  \citep{belloni96, mendez97}.
Some of the intermediate spectra have steep power law tails with $\Gamma>2.4$ leading to
them being called the steep power-law state \citep{mr06}, although
a power law is not a good approximation for the complex curvature of Comptonization 
seen in this state \citep{zdz02,gd03}. This state is 
much more common on the rise than the decline, so is generally a high luminosity state, leading to 
its original name of a very high state (hereafter VHS, \citealt{miyamoto91}). 
Where these spectra are seen on the decline, the disk is always clearly seen as a separate
component from the Compton tail. However, some of the most extreme examples of this
state seen on the rapid rise are simply dominated by the Comptonized component, sometimes
without even a point of inflection in the spectrum to mark the presence of the disk
(e.g. RXTE data of ObsID 30191-01-02-00 of XTE~J$1550-564$ and 
91702-01-58-00 for GRO~J$1655-40$). 

The different states can also be identified by their fast variability properties. In the HSS the 
disk spectrum is remarkably constant on timescales less than a few hundred seconds. This is
expected as the viscous timescale is long  for a geometrically thin, cool disk. Thus 
there is very little variability power at frequencies $f<0.01$~Hz at energies 
where the disk dominates the 
spectrum. Conversely, in the LHS there is strong band limited noise, with a quasi-periodic
oscillation (QPO) superimposed. The frequency of this QPO  termed a type C
increases from 0.1--6~Hz as the
spectrum softens from the LHS into the intermediate states. These intermediate states
can themselves be split into a hard intermediate state (HIMS) and soft intermediate state
(SIMS). As suggested by the nomenclature, the SIMS has stronger disk and weaker
tail than the HIMS, but this is a rather subtle distinction. The main difference 
between these two states is that the HIMS has variability properties which are clearly
similar to those in the LHS, with broadband noise and type C QPO, whereas 
the broadband noise collapses in the SIMS,
leaving the power spectrum dominated by the QPO alone (termed a type B QPO where
this still retains the harmonic structure seen in the HIMS, or a type A QPO when this is
broadened out into a single feature: \citealt{wij99, belloni02,belloni10}). The most extreme VHS spectra noted above
have  power spectra like the HIMS, while those where the disk is clearly visible as 
a separate component most often have power spectra which are similar to the SIMS.

The high energy emission 
in the VHS, HSS, and bright LHS,
is caused by energetic electrons via 
inverse Compton scattering of low energy seed photons from the disk. 
In the HSS, this corona is either optically thin or patchy, so only intercepts a small
fraction of the disk flux, making very little difference to the observed $r_{\rm in}$.
However, in the 
VHS/HIMS and VHS/SIMS,
the corona is clearly both moderately optically
thick and covers most of the inner disk \citep{zdz02,gd03,kd04}. Compton scattering removes photons from the 
disk, boosting them in energy to form the tail. Hence 
the disk appears less luminous, it has an apparently smaller
radius (Kubota et al 2001; Steiner et al 2010). The constant HSS radii can be recovered
by correcting for this effect in spectra where the corona carries less than
half of the accretion power \citep{kubota01,kubota04,steiner10}, but beyond this point 
the inferred radius of the disk is larger than in the HSS, indicating that the disk
may be truncated before the ISCO \citep{kd04, dk06, gdp08}. 
It is essential to understand the precise geometry of 
the optically thick disk and X-ray coronae in these states in order to derive constraints
on the inner disk radius 
and hence test whether the disk begins to recede as predicted by 
the alternative hot accretion flow models.  

In this paper, we utilize Suzaku data of a well known 
stellar black hole GX~$339-4$ in the VHS/HIMS seen during the 2006/2007 outburst 
which is the brightest outburst 
ever observed by RXTE.
In the energy range of 1.5--12~keV, the RXTE/ASM count rate
of peak of the 2006/2007 outburst was $\sim 14\%$ higher than that of 
the 2nd brightest outburst in 2002--2003.
This is one of the best available VHS/HIMS spectra, and hence the 
best to constrain the disk geometry  during the VHS/HIMS in a LHS-HSS transition. 
Several parameters of GX~$339-4$ and its outburst during 2006--2007 are summarized in \S2.
The observation and data reduction
are briefly described in \S3.
The Suzaku spectra are analyzed using
the commonly used {\sc diskbb} plus power-law model (\S4.1), 
two independent corona models (\S4.2), 
and an inner disk-corona coupled model (\S4.3), 
and then, the truncated disk is discussed in \S5.
In \S6, QPO frequency and change of the disk parameters are briefly discussed.
Errors quoted in this paper represent 90\% confidence limit for a single parameter 
unless otherwise specified.

\section{GX~$339-4$ and its X-ray properties}
GX~$339-4$ is a transient Galactic black hole, discovered in the early 1970s~\citep{markert73}.
The system parameters are rather poorly known from optical studies. \cite{hynes04}
give a lower limit of 6~kpc, but their favored distance of 16~kpc is not consistent with the 
measured absorption systems in GX~$339-4$, so a more likely distance is 7--9~kpc \citep{zdz04}. This matches well with models of the binary system, where the low mass companion star
must be big enough to have a high mass transfer rate to power the observed luminosity 
so needs to be substantially more distant than 6~kpc so as not to be seen directly \citep{munos08}.
Hereafter we assume a distance of $8$~kpc.
The orbital inclination is estimated to be less than $75^\circ$ to the line of sight, as 
inferred from the lack of eclipses and dips \citep{cowley02,mari10}. The lack of strong
iron absorption lines in the HSS also argues for $i<~60^\circ$ \citep{ponti12}. 
Conversely there
are lower limits on the inclination from studies of the
binary ($i>45^\circ$: \citealt{hynes03}) while the strong QPO seen in this source also argues
for $i>~50^\circ$ if this is made from Lense-Thirring (vertical relativistic) precession \citep{ingram09}.
We use
$i= 50^\circ$ as it fills all these constraints and is the result obtained 
from the iron line profile analysis of \cite{shidatsu11_low}.

The HSS spectrum of this source was for the first time obtained with {\it Tenma}, 
and \citet{makishima86} estimated the apparent inner radius. With our system 
parameters as used above this corresponds to 
$r_{\rm in}\sim 52\cdot \zeta_{50}d_8$~km,
where $\zeta_{50}$ is $(\cos i /\cos 50^\circ)^{-1/2}$ with the inclination angle $i$, and
 $d_8$ is the distance to the source in the unit of 8~kpc.
 Similar results  are found by \cite{shidatsu11_high}
based on the 2--9 keV MAXI and the 0.6--7 keV Swift data during the 2010 outburst.
The RXTE study of 
\cite{mari10} with multiple datasets gives a larger estimate, 
$r_{\rm in}$ as $67.5\pm0.9 \cdot \zeta_{50}d_8$~km. 
Changing the bandpass to the lower energy range of CCD data gives a
slightly different normalization as the {\sc diskbb} model does not
completely describe the spectral shape of the disk \citep{mari11}. The three disk dominated spectra in XMM-Newton
give $r_{\rm in}=57\cdot \zeta_{50}d_8$~km \citep{mari11}.
Thus the value of apparent inner radius in the HSS is estimated as 
$r_{\rm in}\simeq (52\sim 68)\cdot \zeta_{50}d_8$~km. 
Folding in correction factors of color temperature, $\kappa$, and inner boundary 
condition, $\xi$, gives a 
radius for the ISCO of
$R_{\rm ISCO}=r_{\rm in} \cdot \kappa^2\cdot \xi\simeq (56\sim 73)\cdot \zeta_{50}d_8$~km,
taking $\kappa\simeq 1.7$ \citep{shimura95} and 
$\xi=0.37$ \citep{gier99}. The alternative approach of fitting proper radiative transfer disk
models with full relativistic corrections also gives similar values \citep{mari10}.
 
The disk structure in the LHS has been analyzed in detail by \citet{shidatsu11_low}. 
They  used Suzaku data of this source obtained in 2009 March, and showed that 
the iron line features indicated the inner radius of  
$13.3^{+6.4}_{-6.0}~R_g$, 
where $R_g\equiv GM/c^2$ is the gravitational radius.
It corresponds to $140^{+70}_{-63}~(M/7M_\odot )$~km.
They also showed that weak continuum emission from 
the optically thick disk and surrounding coronae was consistent with the truncated disk 
picture with $R_{\rm in}\sim 120 \cdot \zeta_{50}d_8$~{\rm km}, though again we caution
that this is much more model dependent than the HSS. 

\section{Observation and Data reduction}

\begin{figure}
\epsscale{.80}
\plotone{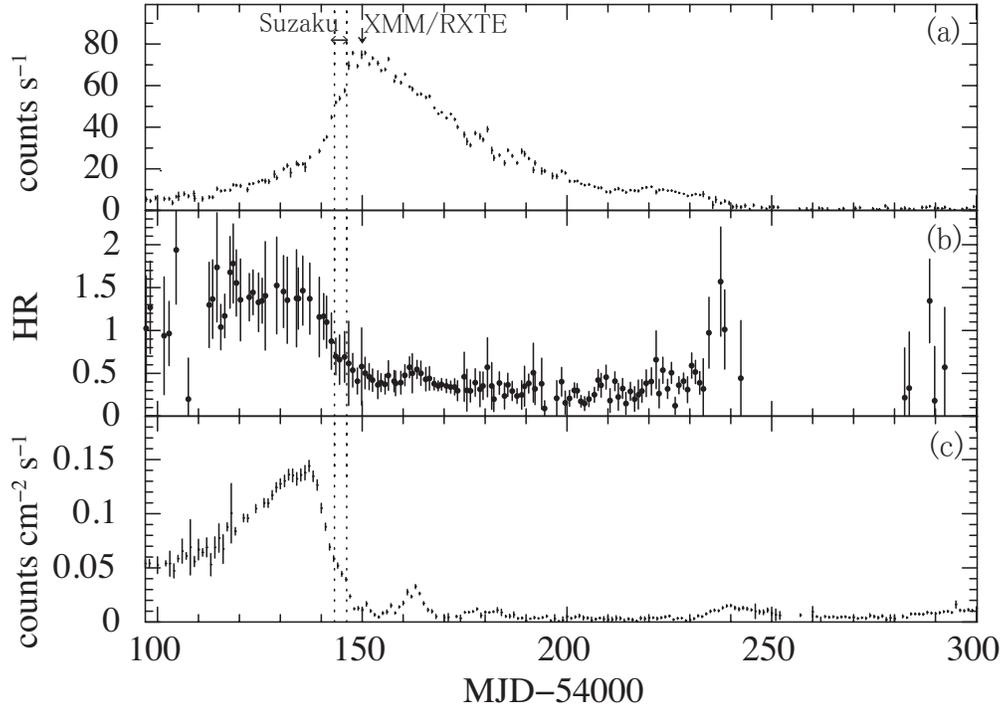}
  \caption{Light curves of GX~$339-4$ based on the RXTE/ASM and the Swift/BAT Hard X-ray Transient Monitor. Panel(a) shows the 1.5--12~keV ASM count rate, 
  while panel(b) shows the ASM hardness ratio (5--12~keV/3--5~keV) where data points of large statistic error ($\Delta {\rm HR} >0.7$) are excluded. Panel(c) shows the 15--50~keV BAT count rate.
A left-right arrow with vertical dashed lines indicates the time period
during which the presented Suzaku data were obtained.  Simultaneous observations of  XMM and 
RXTE on 2007 February 19th is indicated with a downarrow. }
  \label{fig:asm}
\end{figure}

As described in \S1, the 2006/2007 outburst of GX~$339-4$ 
was its brightest outburst since the launch of RXTE in 1995 \citep{swank06}.
Figure~\ref{fig:asm} shows the 1.5--12~keV and 15--50~keV light curve and hardness ratio of 
this outburst, obtained with the RXTE/ASM and 
the Swift/BAT Transient Monitor.
The time histories of the hardness ratio and the Swift/BAT count rate suggests 
that the state transition occurred around MJD$\simeq$54140--54150.
The Suzaku observation was performed on 2007 February 12th 05:33:31 to February 15th 04:48:26 
corresponding to MJD=54143.2--54146.2 as indicated with dashed lines in figure~\ref{fig:asm}.
Clearly the source hardness and the BAT count rate changed dramatically around the time of the Suzaku observation.
An analysis of the QPO seen in RXTE showed a change from type-C to type-B on February 16 \citep{motta09}, so the source made a transition from VHS/HIMS to VHS/SIMS  between 
February 15 and 16. The Suzaku spectrum and variability shows this to be a VHS/HIMS spectrum, 
though there is still an inflection marking the separation of disk and Compton tail \citep{miller08,yamada09}.

The XIS employed the
1/4 window option and a burst option (0.3 s for XIS0/XIS1,
and 0.5 s for XIS3). The Hard X-ray Detector (HXD; \citealt{kokubun07})
was operated in the standard mode. The
data processing and reduction were performed in the same way
as 
\cite{yamada09},
using the Suzaku pipeline processing version 2.0.6.13.
To extract light curves and spectra, the data were analyzed with HEAsoft 
version 6.9 and the calibration data files (CALDB) released on 2007 July 10th.
The XIS events were extracted from a circular region with a radius of 7$^\prime$
centered on image peak.
Since this extraction circle is larger than the window size of $17^\prime  .8\times 4^\prime .5$, the effective extraction
region is therefore the intersection of the window and this circle.
As reported in detail by \citet{yamada09}, the XIS events piled up significantly, 
and thus a central region of $3^\prime$ at the image center was excluded from the
event extraction region to minimize these effects.
Following \citet{yamada09}, we use data from XIS0, since data from XIS1 and XIS3 
are affected more by pile up than XIS0. 
In addition, a variable fraction of the CCD frame was often lost due to
telemetry saturation, resulting in only 2.83~ks of data out of the 
12.4~ks of XIS0 exposure.
While \citet{yamada09} included data affected by the telemetry saturation
to describe the iron line in detail,
we excluded the telemetry saturation to avoid a small effect on the continuum shape
due to the asymmetric response. 

The PIN and GSO spectra, acquired for a net
exposure of 87.9~ks, 
were corrected for small dead time, but no
other correction due to the source brightness was necessary.
We subtracted modeled non-X-ray backgrounds (NXBs; \citealt{fukazawa09}). 
The PIN and GSO background spectra were 
constructed based on {\sc lcfitdt} and {\sc lcfit} 
methods (2.0ver0804), respectively,  filtered by the same good time intervals. 
As the response files, 
{\sc ae\_hxd\_pinxinome3\_20080129.rsp} was used for PIN,
and {\sc ae\_hxd\_gsoxinom\_20080129.rsp} and 
{\sc ae\_hxd\_gsoxinom\_crab\_20070502.arf} were used for GSO. 
The cosmic X-ray background was ignored, since it is less than 1\% of the total counts.
We use GSO data only up to  200~keV,
since the signal becomes smaller than 3\% of the background at 200~keV.
Figure~\ref{fig:suzaku_lc} shows the Suzaku light curves of GX~$339-4$
in which background levels were 
subtracted and dead times were corrected for PIN and GSO data.


\section{Analyses of averaged spectrum}

In this section, we present the analysis of the Suzaku spectra extracted above.
To fit the observed spectra, 
we use the {\sc xspec} spectral fitting package(version 12.6.0). 
We use energy range 0.7--9.0~keV for XIS0,
13--60~keV for the PIN spectrum, and 70--200~keV for the GSO spectrum.
Large fit residuals due to calibration uncertainties are 
often observed near the edge structures of the XIS/XRT instrumental responses so we 
exclude XIS data from 1.4--2.3~keV. We fit the three spectra simultaneously, with constant
factors scaling between the three fixed at  1, 1.07, and 1.07 for 
XIS, PIN and GSO, respectively\footnote{http://heasarc.gsfc.nasa.gov/docs/suzaku/analysis/watchout.html}.
We extend the energy range of the spectral fitting to 0.1--1000~keV as we use some
convolution models. Such models have edge effects at the end of the 
energy range used for the calculation, so it is important that this beyond the energy range used
for the data.

\subsection{Empirical modeling by {\sc diskbb} and {\sc power-law} with reflection (model~1)}

In order to characterize the spectral shape, we first model the 0.7--200~keV Suzaku data 
in the same way 
as \citet{yamada09}, i.e.,  the {\sc diskbb} model plus power-law 
(hereafter PL) model. 
We hereafter call this `model~1'. The two continuum components are both absorbed by a 
common hydrogen column modeled with {\sc xspec} {\sc wabs} model~\citep{morrison83}.
While \citet{yamada09} used {\sc laor} model to describe the iron line in detail,
we use a single gaussian with $\sigma$ at 0.2~keV
since we focus on the continuum shape and our much shorter exposure (due to more stringent 
constraints on telemetry issues)  means that the
spectra have much lower statistics. 

To account for ionized reflection of the PL component, we utilized the  model
{\sc ireflect} (a convolution version of the {\sc pexriv model} \citealt{mz95}).
This model balances photoionisation with radiative recombination
from the very simplistic ionization balance code of \citep{done92}.
This requires a user defined  (rather than self-consistently computed) 
temperature, which we set to $10^6$~K as the disk is clearly hot. 
However, 
the code does not calculate collisional processes,
so the ion population are set only  by photoionisation. This is clearly
a poor assumption for such a hot disk.
Thus the fairly 
high ionisation parameter 
 required by \cite{yamada09}
may be an overestimate of the illuminating flux required. Yet another limitation is that 
Compton scattering within the disk is not included, which has quite a large effect on the
characteristic iron line and edge profile in the reflected emission. 
 Photo-ionization heats the
disk photosphere to some fraction of the Compton temperature, so this can be of order 
 $\sim 1$~keV for high ionization parameters. All the line/edge features are produced at
an optical depth of around unity, so this means that a third of the photons are Compton 
scattered by these hot electrons, smearing the sharp line/edge seen in 
{\sc ireflect/pexriv}  \citep{ross99, ross05}. This may mean that the relativistic smearing is overestimated, but
we caution that generally spectral parameters are all interrelated, so it may
instead distort the observed solid angle, ionization state or iron abundance. 
Despite these drawbacks, we use this as there are few better 
alternatives at present. 
The models at high densities of \cite{ross07} are not yet public.

We fix the ionization parameter $300~{\rm erg~cm~s^{-1}}$, and fix all abundances 
other than iron at solar. Hence the only free parameters are the amount of solid angle 
and the iron abundance. We smear this reflected spectrum by  general relativistic effects
using {\sc rdblur}~\citep{fabian89}.
We fix all the parameters of this at characteristic values seen by \cite{yamada09}, 
i.e. inner and outer radius 
$R_{\rm in} ^{\rm rdblur}=10R_g$, $R_{\rm out} ^{\rm rdblur}=10^5R_g$, 
inclination angle $i=50^\circ$, and power law index of emissivity $\beta=-3$ 
as our
limited data quality do not allow us to constrain them independently. 

\begin{figure}
\epsscale{.80}
\plotone{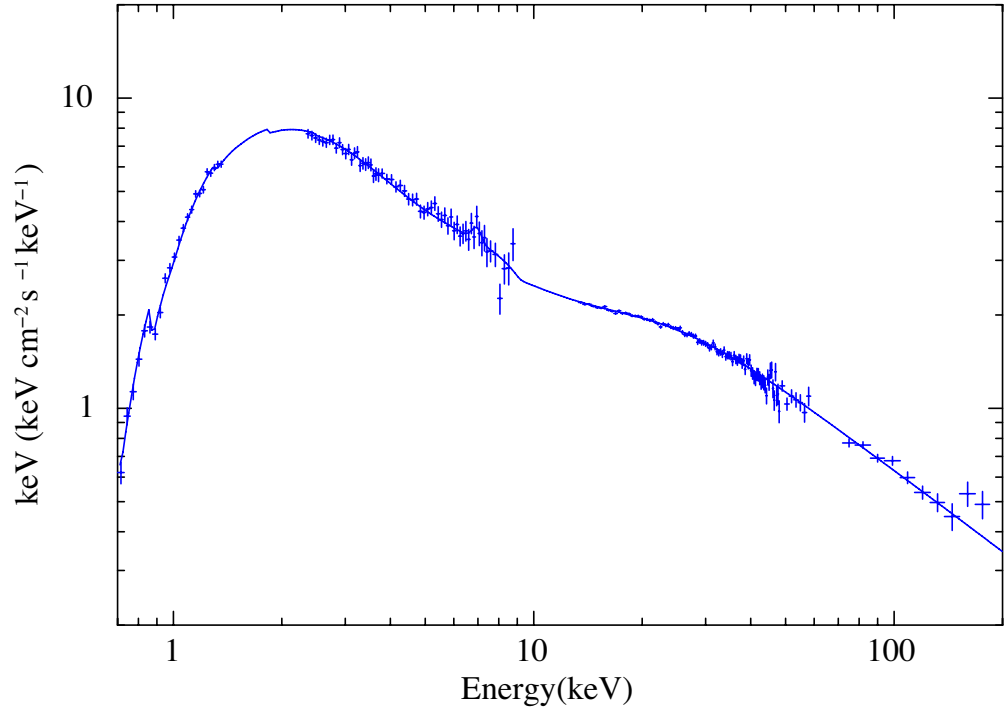}
   \caption{$\nu F_\nu$ spectrum of GX$339-4$ from the best fit  model~1, where PIN and GSO spectra are normaized to XIS-0 spectrum. }
  \label{fig:eeu}
\end{figure}

\begin{figure}
\epsscale{1.}
\plottwo{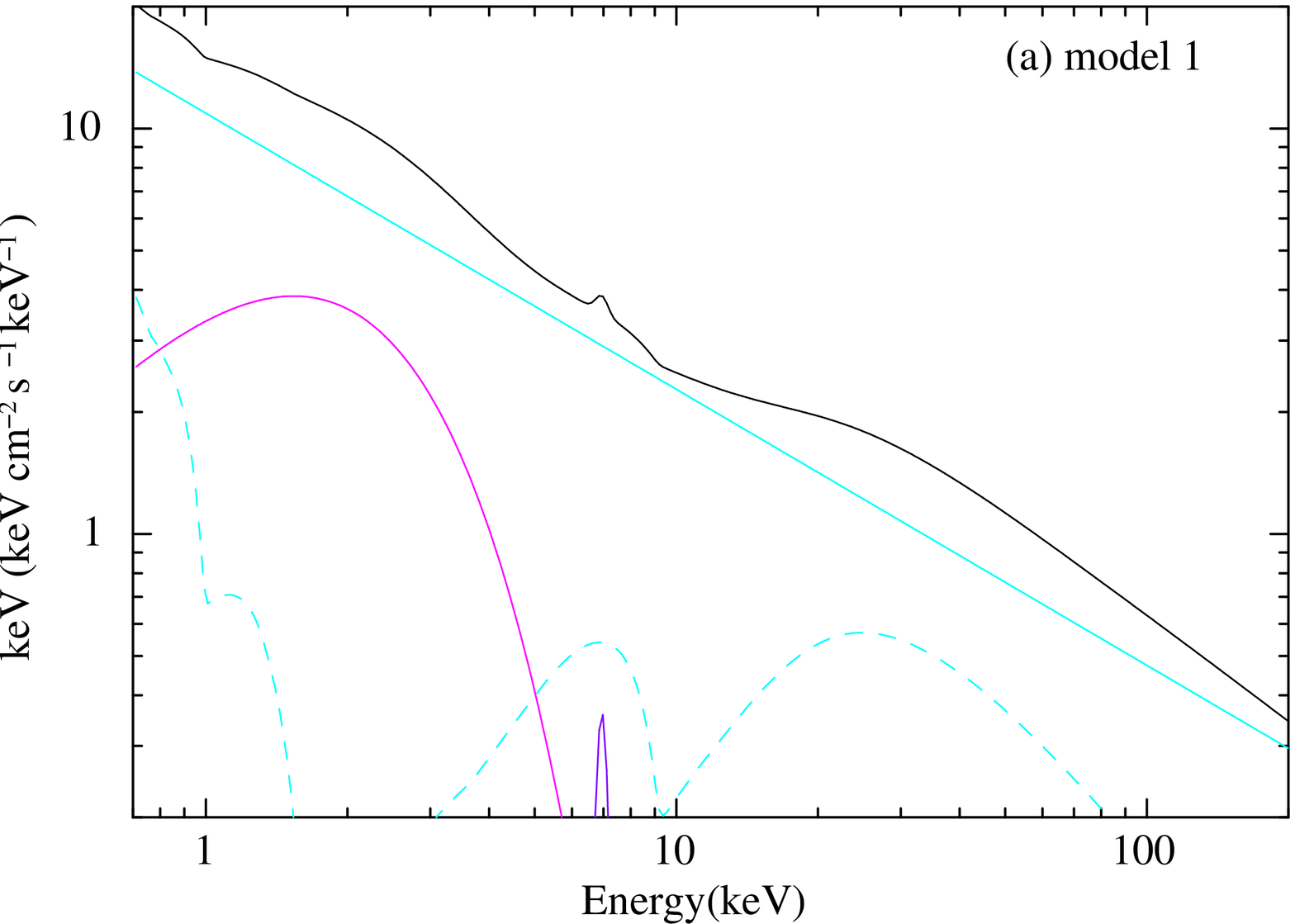}{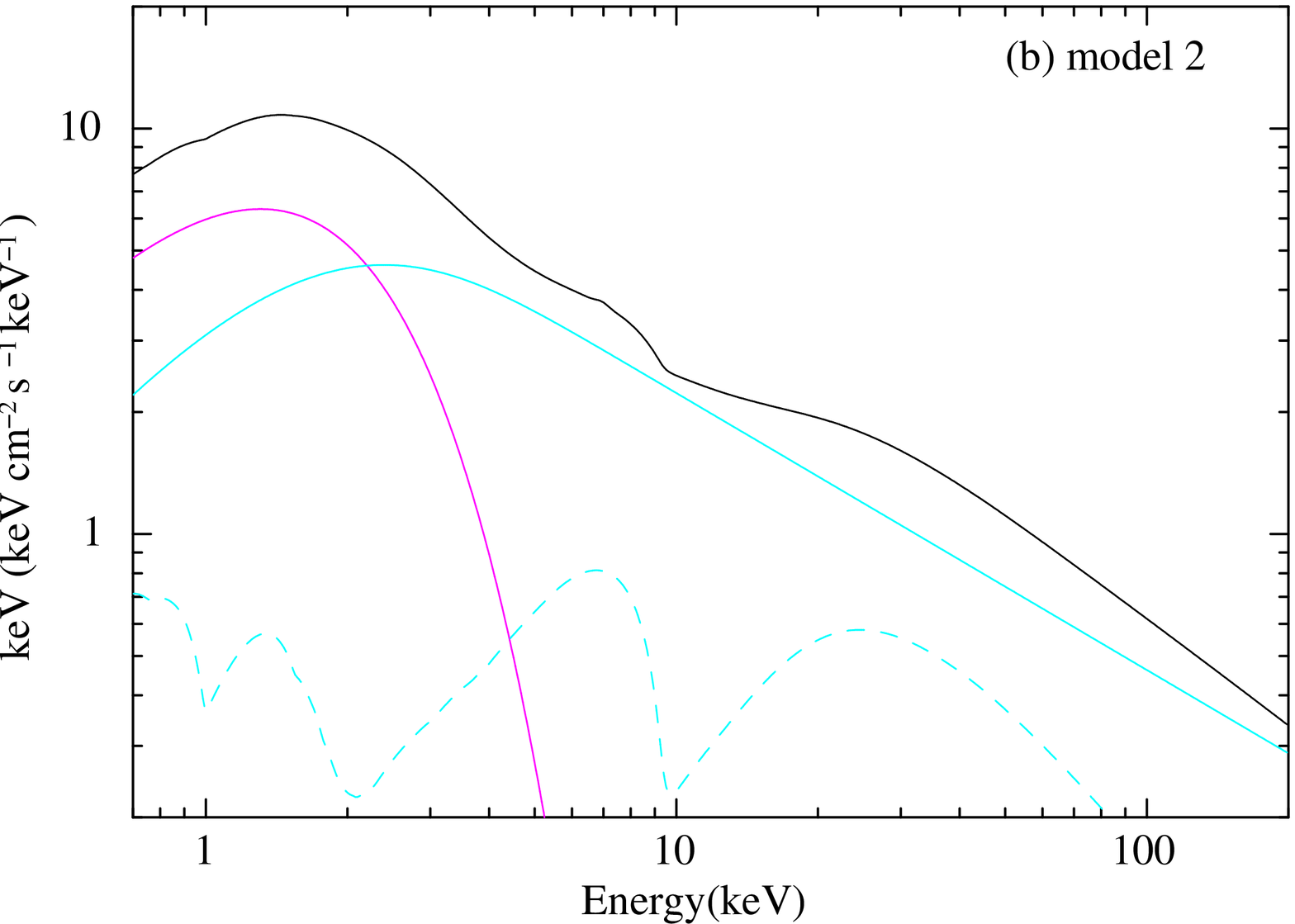}

\plottwo{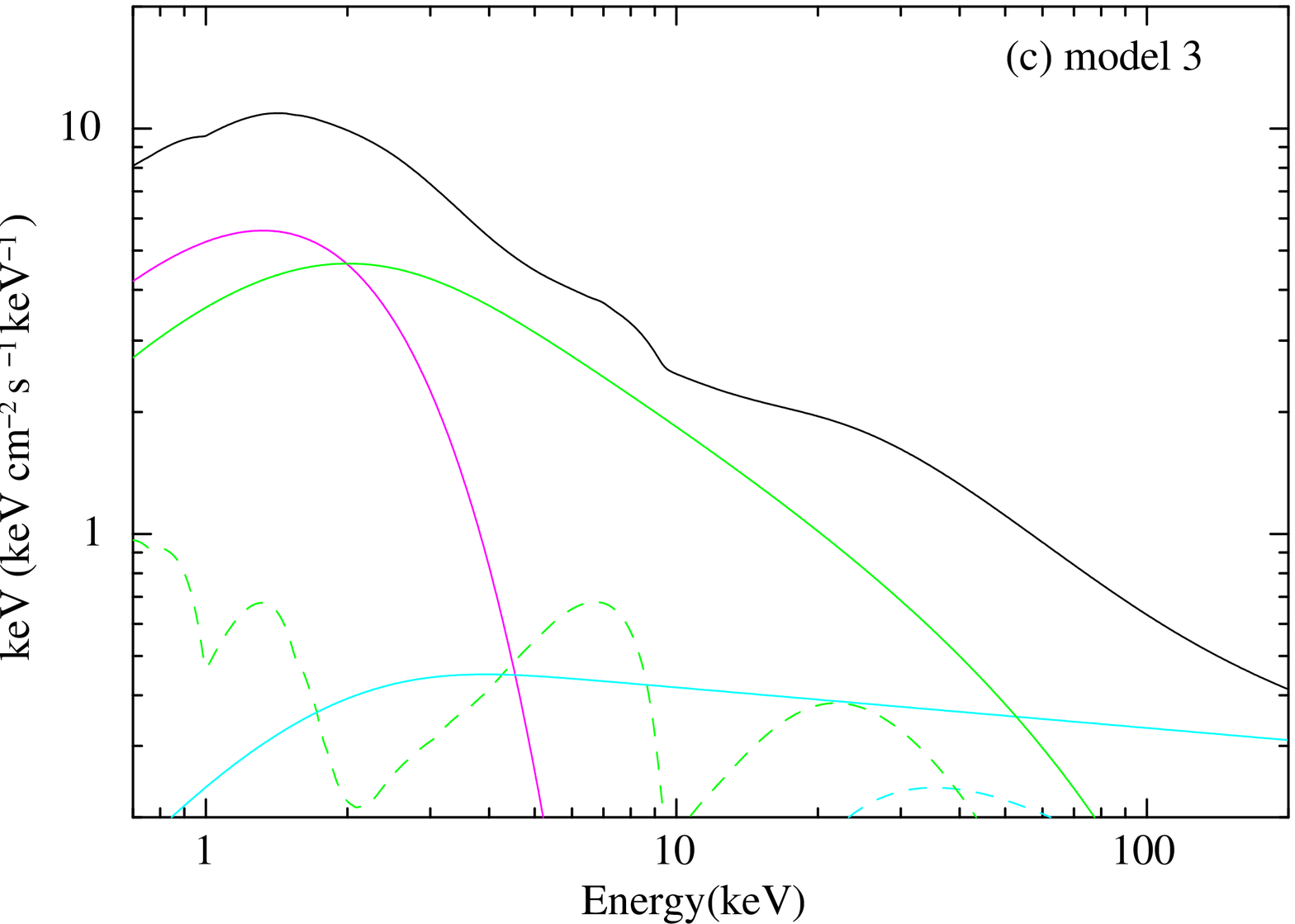}{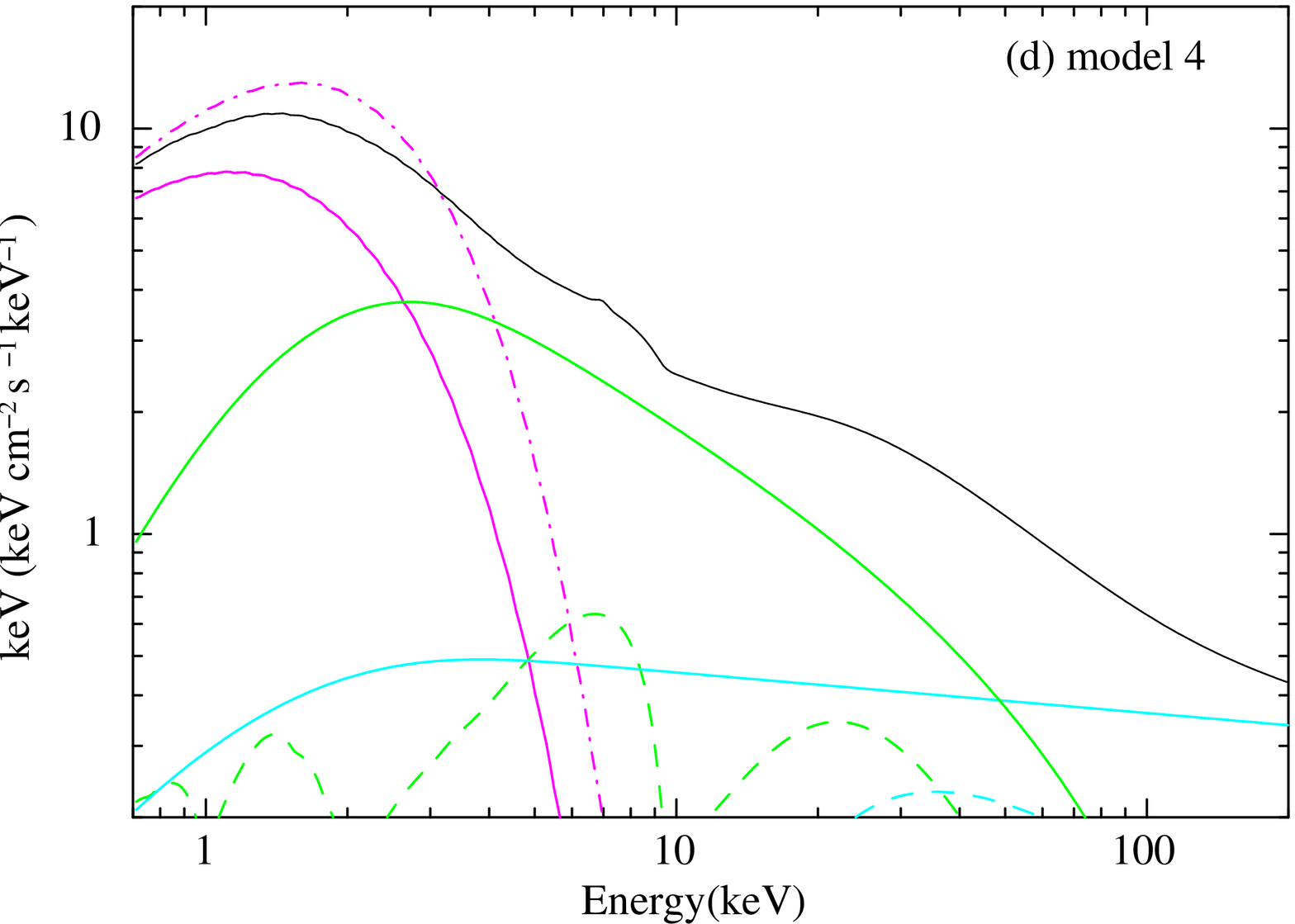}
   \caption{Model spectra of GX$339-4$ from the best fit  model~1~(panel~a), model~2~(panel~b), model~3~(panel~c),
   and model~4~(panel~d), where the interstellar absorption is excluded. Emission from 
   the underlying disk ({\sc diskbb} or {\sc dkbbfth(d)}; magenta), non-thermal compton component ({\sc pl} or {\sc simpl(c)}; light blue), 
   thermal Compton component ({\sc nthcomp} or {\sc dkbbfth(c)}; green), and gaussian (purple) are shown solid lines.  The reflected components of the thermal (green) and 
   non-thermal (light blue) are shown with dashed lines. In panel (d), the intrinsic disk emission (magenta dash-dot  line),
    where all the power of the {\sc dkbbfth} component is released as optically thick disk, is plotted
     together with the original model components.
   }
  \label{fig:eemo}
\end{figure}

Table~\ref{tab:models} gives details how the model was described in {\sc xspec}, while 
Table~\ref{tab:bestfit_sum} gives the best fit parameters. This shows 
that model~1 fits the data well with $\chi^2/dof=190.2/186$.
Figure~\ref{fig:ld-delch} shows the data and the best fit model and 
residuals, while figure~\ref{fig:eeu} and figure~\ref{fig:eemo}(a) show the
$\nu F_\nu$ spectrum and unabsorbed model spectrum, respectively.
The absorbed 0.7--200~keV flux is estimated as 
$2.8\times 10^{-8}~{\rm erg~s^{-1}cm^{-2}}$ 
which gives an absorbed luminosity of
$2.1\times 10^{38}~d_8~{\rm erg~s^{-1}}$ assuming isotropic emission.
The PL photon index $\Gamma_{\rm PL}=2.68^{+0.02}_{-0.01}$, so this plus the high luminosity
means that the source is in the VHS, as noted by \citet{yamada09}.
The best fit parameters are consistent with those obtained by \citet{yamada09} within 90\% confidence, and 
the best fit {\sc diskbb} model shows disk inner temperature of 
$kT_{\rm in}=0.65\pm 0.02$~keV and an apparent inner radius of
$r_{\rm in}\sim (56^{+5}_{-6})\cdot \zeta_{50}d_8$~km, consistent with that observed in the 
usual HSS, $(52\sim68)\cdot \zeta_{50}d_8~{\rm km}$ (see \S2). 

However, an inspection of the figure~\ref{fig:eemo}(a) shows
that this is not a good physical description of a model where the disk
provides the seed photons for Compton up-scattering into the PL
tail as the PL extends below the disk at low energies. 
This motivates further studies of the disk structure with more physical models.


\subsection{Disk geometry based on independent corona (model~2 and model~3)}

\begin{figure}
\epsscale{.80}
\plotone{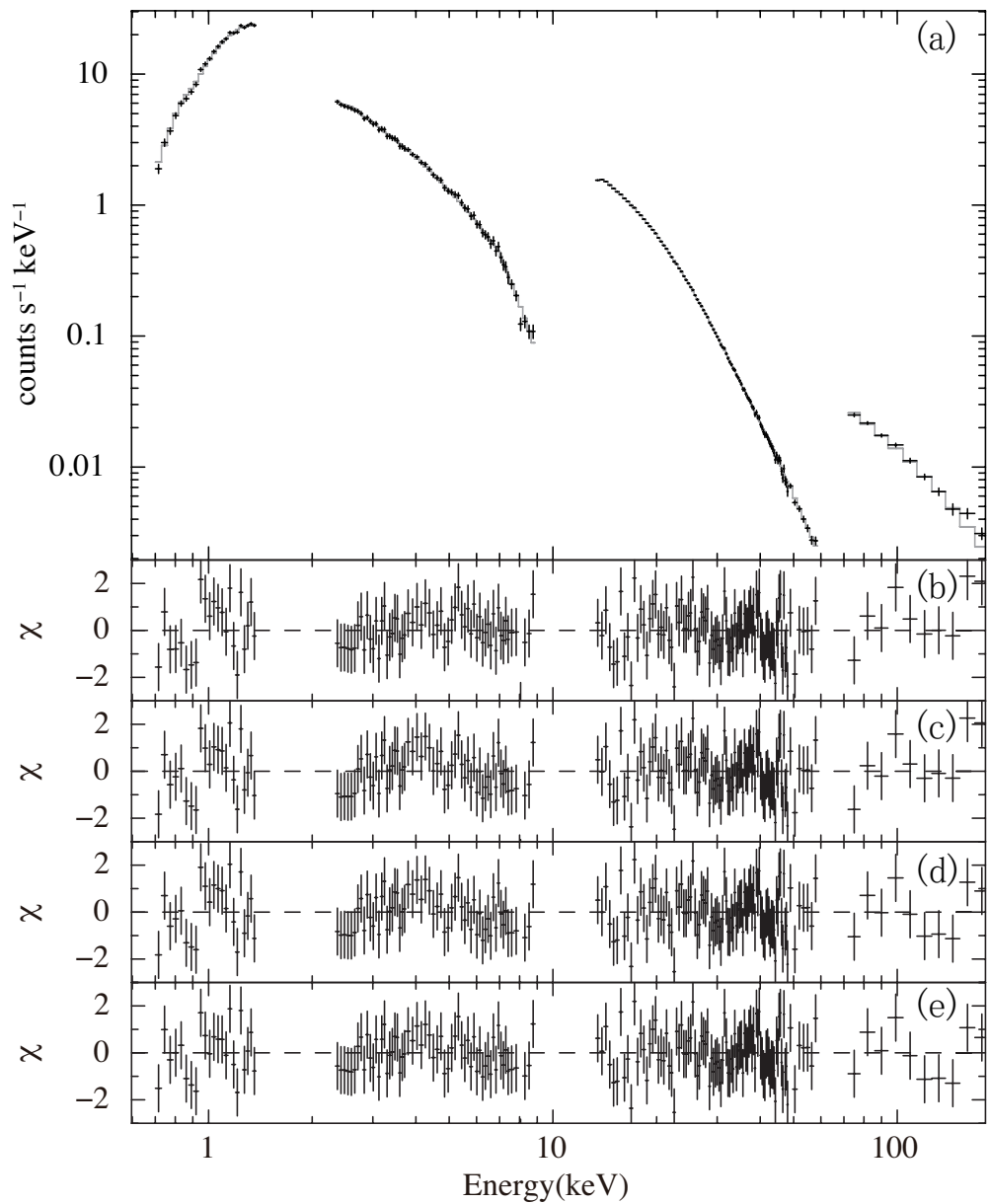}
   \caption{The Suzaku raw spectra of GX~$339-4$, together with predictions 
	  of the best-fit model~1 are shown in panel~(a).  
   Residuals between the data and model~1,
 model~2, model~3, 
 and model~4, are shown in panel~(b), (c), (d), and (e), respectively.
}
  \label{fig:ld-delch}
\end{figure}

Compton scattering conserves photon number, so it removes seed photons from the observed
spectrum by boosting them
in energy to form the Compton tail. We replaced the 
{\sc pl} model with a convolution model {\sc simpl} \citep{simpl} 
which self consistently removes as many seed
photons as it puts into a Compton tail.
This is identified as `model~2' in  table~\ref{tab:models}. 
As described in table~\ref{tab:models},  
only the up-scattered photons are reflected by the optically thick disk, and thus 
modified by the models of {\sc rdblur} and {\sc ireflect}.
The fit results are presented in table~\ref{tab:bestfit_sum} and figure~\ref{fig:ld-delch}(c). 
As shown in unabsorbed model spectrum (figure~\ref{fig:eemo}b), this gives a more physical 
description of the distribution of the low energy photons. The lack of
emission at the lowest energies from Comptonization compared to a PL means that the
disk temperature becomes lower to compensate for this, with 
$kT_{\rm in}= 0.55^{+0.02}_{-0.01}$~keV, and the radius becomes higher
$r_{\rm in}=(122^{+7}_{-8})\cdot \zeta_{50}d_8$~km.
This implies that the disk inner radius is twice as large as the ISCO, 
by comparison to that found in the usual HSS.

However, the Comptonization seen in the VHS is complex, containing a mix of both thermal
and non-thermal electrons 
(e.g., \citealt{gier99, kubota01, gd03})
whereas the {\sc simpl} model used above
only makes a non-thermal power law tail. Hence we added
a thermal comptonization model, {\sc nthcomp} 
\citep{zdz96,zycki99} to the model (model~3 in table~\ref{tab:models}). 
This has four parameters, seed disk photon temperature, which we
tie to that of the seed photon temperature  of {\sc diskbb}, the 
electron temperature $kT_{\rm e}$, photon index $\Gamma_{\rm th}$ (equivalent to optical 
depth), and normalization. The relation between the optical depth, electron temperature
and spectral index is given by equation (A1) of 
\cite{zdz96} assuming a spherical source 
with a uniform distribution of seed
photons throughout the source. 
By contrast, we assume slab geometry with seed photons 
at the bottom of the slab.
Comparing these two geometries with {\sc compps} (geom=$-4$ and 1 respectively, \citealt{compps})
shows that the spectral indices are equivalent for an optical depth in the slab geometry which is 
approximately half that of the sphere. Hence we use
\begin{equation}
\tau=\frac{1}{2}\cdot \left(\sqrt{\frac{9}{4}+\frac{3}{\Theta_{\rm e}\cdot ((\Gamma_{\rm th}+\frac{1}{2})^2-\frac{9}{4})  }}-\frac{3}{2}\right)~~
\end{equation}
where $\Theta_{\rm e}=kT_{\rm e}/m_{\rm e} c^2$. 

We cannot uniquely constrain both $\Gamma_{\rm th}$ from the thermal
Comptonization, and $\Gamma_{\rm pl}$ from {\sc simpl} so we fix the latter at 
2.1, the value of the non-thermal tail seen in the typical HSS \citep{gier99}.
The non-thermal tail may be somewhat steeper in the VHS, and again has a complex
shape \citep{gd03}. However, our data do not extend above 200~keV 
so we are not very sensitive to this. Hence model 3 has 
only two additional free parameters.
Similarly to models~1 and model~2, only the Comptonized emission 
is reflected by the underlying optically thick disk, 
and thus the model description is somewhat complicated as described in table~\ref{tab:models}.
This gives  $\chi^2/dof=192.0/184$, still not as good as model~1, but better than model~2.
The fits are shown in table~\ref{tab:bestfit_sum}, figure~\ref{fig:ld-delch}(d) and figure \ref{fig:eemo}(c). These show that 
the thermal Comptonization dominates the spectrum below $\sim 50$~keV.

Unlike {\sc simpl}, the thermal Comptonization does not decrease the seed photon
spectrum. Hence we have to manually adjust the disk normalization for the photons which are
scattered out into the thermal Compton tail. Hence we also tabulate the 
unabsorbed {\em photon}
flux $F^{\rm photon} _{\rm th}$ 
contained in the thermal Compton tail integrated from 0.01-100~keV. We assume that 
these photons were removed equally from all energies of the disk spectrum (equivalent to
a uniform corona over the entire disk). 
The observed photon flux from the disk, $F^{\rm photon} _{\rm disk}$ is also tabulated, and we simply increase
the measured disk normalization by this amount. Since the normalization is
proportional to the square root of the radius this gives an
inferred radius $r_{\rm in}^\ast=r_{\rm in} \sqrt{(F^{\rm photon} _{\rm th}+F^{\rm photon} _{\rm disk})/F^{\rm photon} _{\rm disk}}$, where $r_{\rm in}$ is calculated from the {\sc diskbb} normalization alone. This  is equivalent to equation (A.1) in \cite{kubota04},
and this gives
$r_{\rm in}^\ast=(117\pm 5) \cdot \zeta_{50}d_8$~km, 
which is as large as that estimated with model~2.

\subsection{Inner disk-corona: coupled energetics (model~4)}

\begin{figure}
\epsscale{0.8}
\plotone{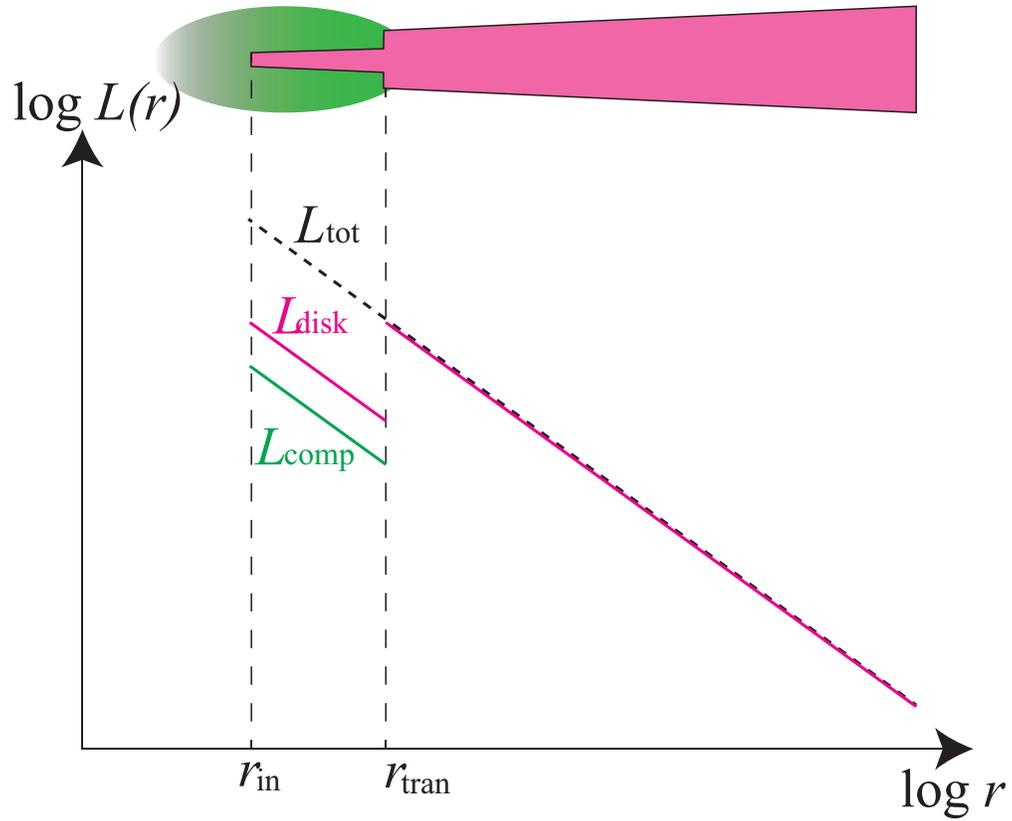}
  \caption{The schematic geometry envisaged for the {\sc dkbbfth} model. 
  At $r<r_{\rm tran}$, sum of the local luminosity from thermal corona (green) and disk (magenta) are consistent with that expected from the standard disk at $r>r_{\rm tran}$. Details are described in \citet{dk06}. }
      \label{fig:model}
\end{figure}

The previous two models assumed that 
the Comptonizing electrons fully covered the disk, and that 
the underlying disk structure is unaffected by the presence of the high energy electrons. 
However, both these assumptions are probably not appropriate. The
corona probably only covers the inner disk, at $r_{\rm in}< r<r_{\rm tran}$ as 
schematically shown in figure~\ref{fig:model}. 
The power in the corona must also derive ultimately from the 
accretion flow, so there should be less power dissipated in the disk.

In the view of  these concepts, we fit the data by replacing the '{\sc diskbb+nthcomp}' 
components in model~3 to  the coupled disk-corona model, {\sc dkbbfth} given by 
\citet{dk06} (model~4 in table~\ref{tab:models}). This model assumes that
the energy released by gravity is dissipated locally, either thermalising in the disk
for $r > r_{\rm tran}$, or split between the disk and corona for $r_{\rm in}<r< r_{\rm tran}$,
as shown schematically in figure~\ref{fig:model}. Thus the disk underlying the corona
is cooler and less luminous than it would have been if the corona did not exist, and 
it is this weak and cool disk emission which provides the seed photons for the Compton
upscattering (see \citealt{sz94, dk06}).

The model parameters are similar to that of using a {\sc diskbb+nthcomp} continuum (model~3),
except that rather than having two separate normalizations, the parameters which 
control the relative normalization of the disk and corona are a combination of 
$r_{\rm tran}$, which determines
the outer, unComptonized disk luminosity, and the shape of the thermal Comptonization
i.e. $\Gamma_{\rm th}$ (equivalent to $\tau$) and $kT_{\rm th}$. The luminosity
of the Compton component, $L_{\rm th}$, is approximately determined by the Compton 
$y$ parameter where $y\approx 4\Theta_e (\tau+\tau^2)$ as 
$L_{\rm th}=y L_{\rm disk,in}$  and
$L_{\rm disk,in}$ is the inner disk luminosity for $r_{\rm in}<r<r_{\rm tran}$.
Thus the ratio of power dissipated 
in the corona to that of the disk in the region $r_{\rm in}<r<r_{\rm tran}$ 
is $f_{\rm th}=L_{\rm th}/L_{\rm disk,in}\approx y$
(see  \citealt{dk06} for examples of how
the model spectra change as a function of these parameters). 
The inner radius is calculated via equation (A.1) in 
\cite{kubota04} by replacing `$F_{\rm disk} ^{\rm p}+F_{\rm thc} ^{\rm p} \cdot 2\cos i $'
to photon flux of the {\sc dkbbfth} component,  $F_{\rm dkbbfth} ^{\rm photon}$, assuming the slab geometry.
The radius is also determined from the {\it intrinsic} (rather than observed) innermost disk temperature, 
$T_{\rm in}^{\rm int}$. This is how the entire disk would be seen if there was no energy
dissipated in the corona.

This model only takes into account the thermal Comptonization, so we again include
the {\sc simpl} with fixed $\Gamma_{\rm PL}=2.1$ to describe the non-thermal tail. 
In this analysis, we set the un-scattered component of the {\sc dkbbfth} to be 
seed photons for {\sc simpl}.
Again, both the thermal ({\sc dkbbfth}) and non-thermal ({\sc simpl}) Compton components 
are both modified by the ionized reflection and blurred by the relativistic effects.

The model has the same number of free parameters as model 3, but gives an
improved fit with $\chi^2/dof=178.7/184$. This is the best chi-squared value among the four
models despite the fact that the model is more constrained.
The fit results are summarized in table~\ref{tab:bestfit_sum}, figure~\ref{fig:ld-delch}(e) and 
figure~\ref{fig:eemo}(d). 
With the best fit parameters by assuming an isotropic emission, the 
absorbed and unabsorbed luminosity was estimated in the range of 0.7--200~keV as
$(2.1\times 10^{38})d_8~{\rm erg~s^{-1}}$ and $(2.9\times 10^{38})d_8~{\rm erg~s^{-1}}$, 
respectively. We also calculated the 
bolometric luminosity 
$(3.8\times 10^{38})d_8~{\rm erg~s^{-1}}$ and the
optical depth of $\tau=0.58^{+0.08}_{-0.07}$ via equation~(1).
The corona was found to be localized within  $r_{\rm tran}$ of  $2.7~r_{\rm in} ^*$, 
within which, 27\% of the accretion power is dissipated in the thermal corona.

Considering the power dissipated in the corona, the intrinsic disk temperature 
should be $kT_{\rm in}^{\rm int}=0.67^{+0.04}_{-0.08}$~keV.
With this intrinsic disk temperature and the 0.1--100~keV {\sc dkbbfth} photon flux of 
$F_{\rm dkbbfth} ^{\rm photon}=43.5~{\rm photons ~s^{-1}cm^{-2}}$, 
the underlying disk inner radius
was estimated as 
$r_{\rm in}^\ast =(93^{+19}_{-8}) \cdot \zeta_{50}d_8$~km.
In figure~\ref{fig:eemo}(d), 
the intrinsic disk emission is plotted together 
with a dash-dot line.
Though the estimated apparent inner radius 
is slightly smaller than that obtained under the independent corona modeling (model~2 and 3), 
it is $\sim1.3$--2.2 times as large as that observed in the HSS (see \S2).

We checked that the results were not affected by the assumption that
relativistic blurring has a radial dependence parameterized by  $\beta=-3$. 
We replaced this by $\beta = 10$, which is more appropriate for a
stress-free inner boundary condition by assuming that the line
emissivity is proportinal to the local energy release rate in the disk.
The difference from the $\beta = -3$ case was negligible.
Even removing relativistic effects entirely does not change the continuum parameters
significantly due to the limited statistics around the iron line/edge in our data. In all cases
the best fit disk parameters 
were still consistent with those shown in table~\ref{tab:bestfit_sum}.


\section{Is the disk truncated in these data? }

\begin{figure}
\epsscale{1.0}
\plottwo{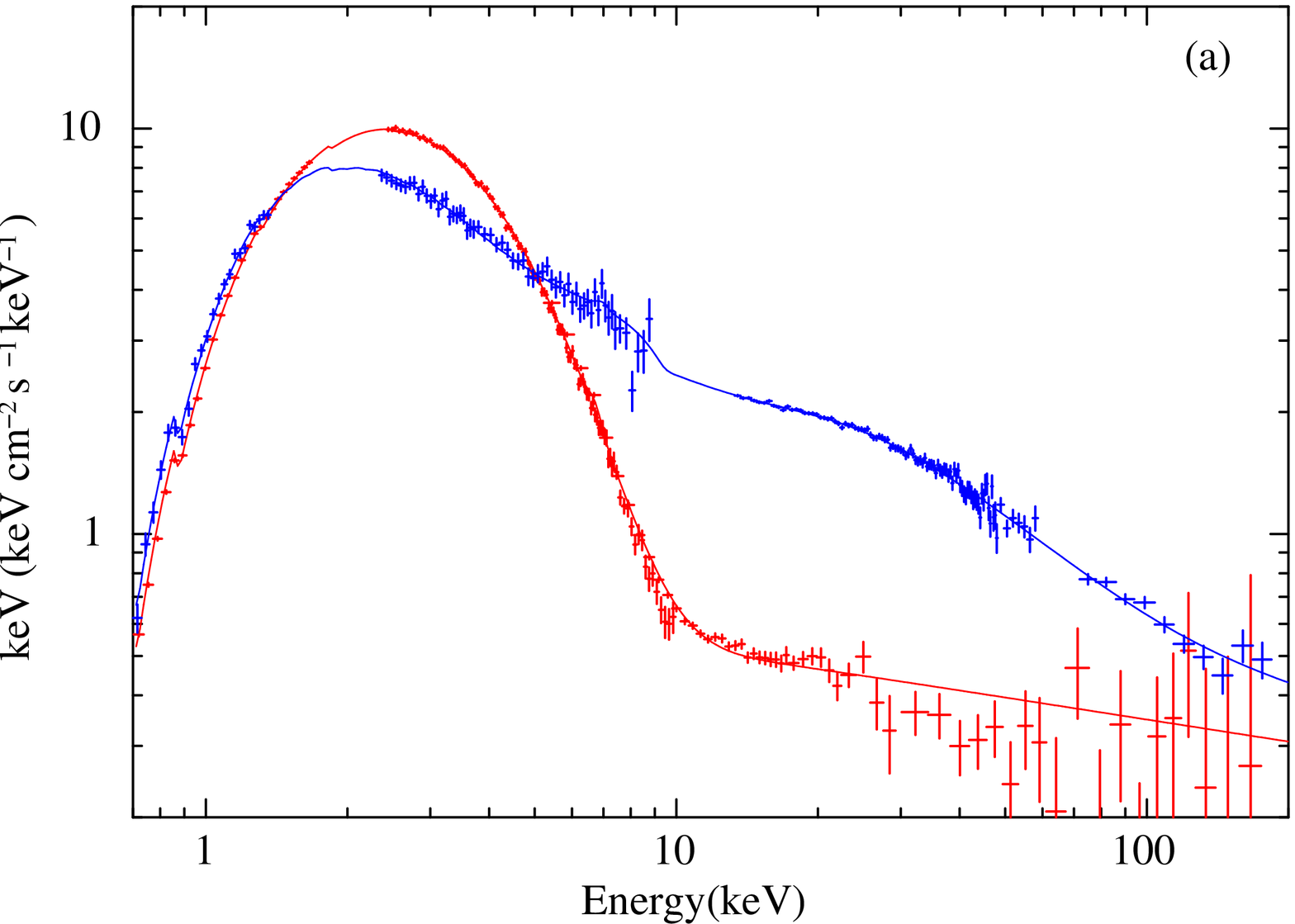}{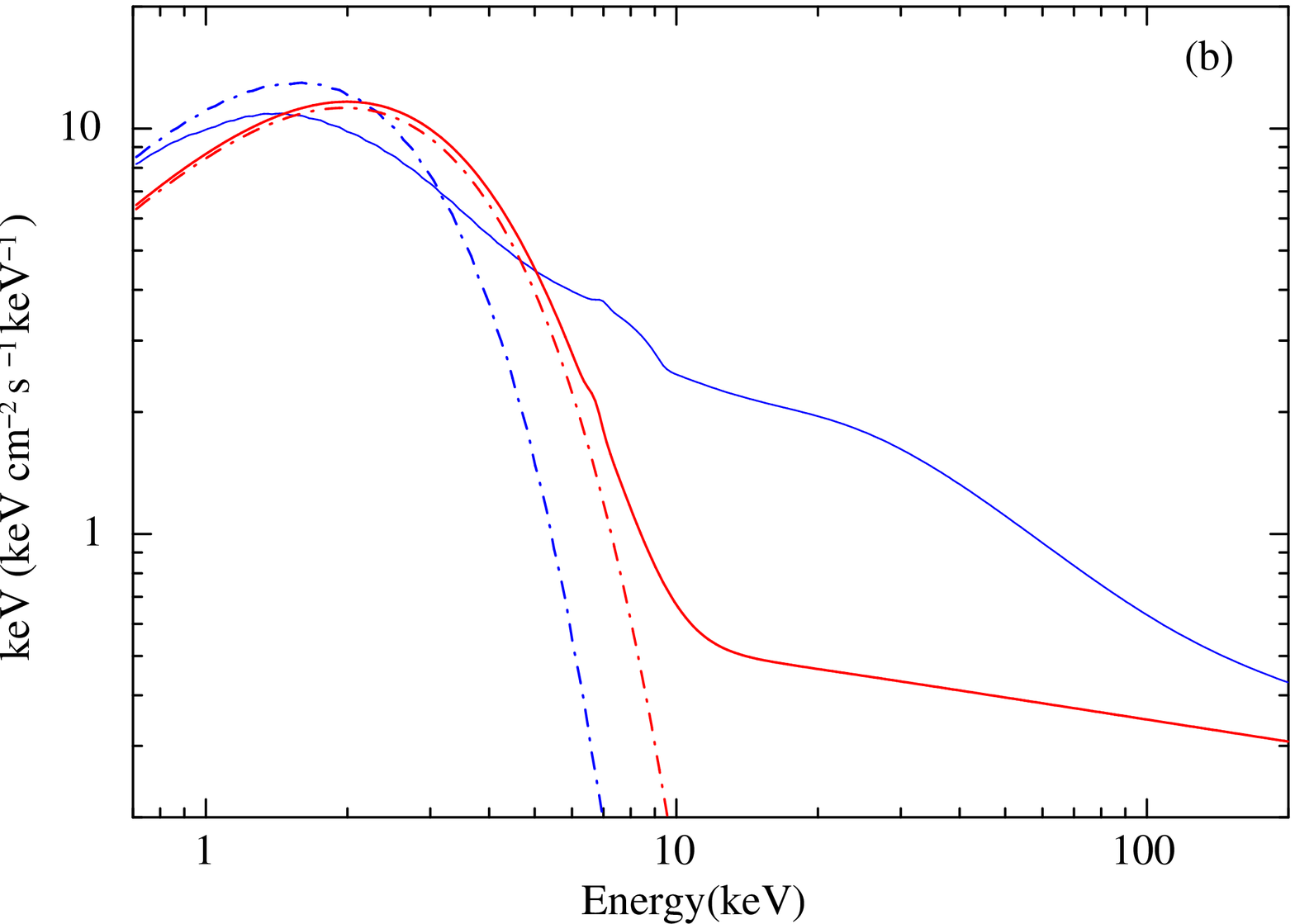}
  \caption{Comparison between the Suzaku averaged spectrum in the VHS/HIMS (blue) and XMM/RXTE spectrum in the HSS obtained on February 19 (red).  $\nu F_\nu$ spectra and the unabsorbed model spectra by excluding the same interstellar absorption are shown in panel (a) and (b), respectively.  
 A blue dash-dot line indicates the intrinsic disk emission for the Suzaku VHS spectrum in panel~(b). The disk emission in the HSS is also indicated with a red dash-dot line.}
  \label{fig:hikaku}
\end{figure}

All the models above which self consistently correct the disk normalization for Compton scattering
(models 2, 3 and 4)
find a disk inner radius which is larger than that seen in the HSS in this source. 
This can be seen directly by comparison with a HSS spectrum at a very similar luminosity.
Figure~\ref{fig:hikaku}(a) shows the Suzaku data used here together with the 
HSS spectrum of almost the same luminosity taken by
XMM and RXTE on February 19th \citep{mari11}
(see figure~\ref{fig:asm}).
Figure~\ref{fig:hikaku}(b) shows the comparison of the unabsorbed model spectra.

The HSS spectrum is roughly 
characterized by a dominant {\sc diskbb} of $kT_{\rm in}\simeq 0.82$~keV with a weak PL tail, and 
an absorbed 0.7--200~keV flux is estimated as  $2.4\times 10^{-8}~{\rm erg~s^{-1}cm^{-2}}$, 
which is only 15\% lower than the Suzaku flux 
of $2.8\times 10^{-8}~{\rm erg~s^{-1}cm^{-2}}$ in the same 
energy band. 
Thus if the disk has constant radius between the HSS and VHS/HIMS then the disk temperature should be
slightly higher in the higher luminosity VHS/HIMS. 
Yet is it clear from this figure that the VHS disk has a lower temperature, as the data peak at 
lower energy in figure~\ref{fig:hikaku}(a). Compton scattering retains the imprint of the seed
photon energy, so this lower temperature cannot be a consequence of simply Comptonizing the
higher temperature inner disk (see \citealt{dk06}). The blue dash-dot line in 
figure~\ref{fig:hikaku}(b) shows the estimated 
intrinsic disk emission from the VHS/HIMS data as reconstructed from the {\sc dkbbfth} model,
i.e. how the disk emission would have looked in the Suzaku data, 
if the thermal corona was not present
and the matters of thermal corona were all accreted in the disk.
If the inner radius is kept constant between the HSS and the VHS/HIMS, 
the intrinsic disk emission should peak at higher energy than  in the HSS. 
However, the peak energy of the intrinsic disk emission is still much lower than that in the HSS.
Estimated intrinsic VHS disk temperature 
$kT_{\rm in} ^{\rm int}= 0.67^{+0.04}_{-0.08}$~keV is much 
lower than the 
observed HSS disk  temperature of $kT_{\rm in}\simeq 0.82$~keV.
Therefore, the apparent inner radius of the VHS/HIMS 
is clearly larger than that in the HSS. 

To convert from apparent to true inner disk radius 
requires corrections for the
color temperature, $\kappa$, and inner boundary condition, $\xi$,  
with $R_{\rm in}= r_{\rm in} \cdot \kappa^2 \cdot \xi$ as 
described in \S2. 
If we use the same $\kappa$ and $\xi$, we obtained $R_{\rm in}=(99^{+20}_{-8})\cdot \zeta_{50}d_8$~km.
Under the same value of $\xi$,
the only way to get the same  inner radius of the disk 
in the VHS/HIMS as in the HSS is if $\kappa$ is as low as $\sim1.3$--1.4. This
seems very unlikely given the strong illumination. 
On the contrary, it  is certainly plausible that the disk color temperature correction has
increased substantially, as the inner disk should be strongly illuminated by the Comptonized 
emission in the VHS/HIMS whereas it is not in the HSS. 
This only reinforces the change in true inner radius as a larger color temperature correction means that
$R_{\rm in}$ is even larger than derived from assuming $\kappa=1.7$. Similarly, if the disk in 
truncated, then the use of the same boundary condition is not appropriate as it now does
not extend down to the ISCO, where there is the stress-free inner boundary, but is truncated
where there can still be stress on the inner boundary, making $\xi$ larger. 
This again leads to an increase in $R_{\rm in}$ (see also the discussion in \citep{gdp08}.

Thus both of the expected changes in color temperature and inner boundary condition 
reinforce the conclusion that the disk inner radius is larger in the VHS/HIMS than in the HSS. 
Using $\kappa=1.9$ \citep{shimura95} and $\xi=1$ as a reference, 
$R_{\rm in}$ can be as large  as 
$\sim 340\cdot \zeta_{50}d_8$~km. 

\section{Variability of the disk and corona}

\subsection{QPO frequency}

\begin{figure}
\epsscale{.80}
\plotone{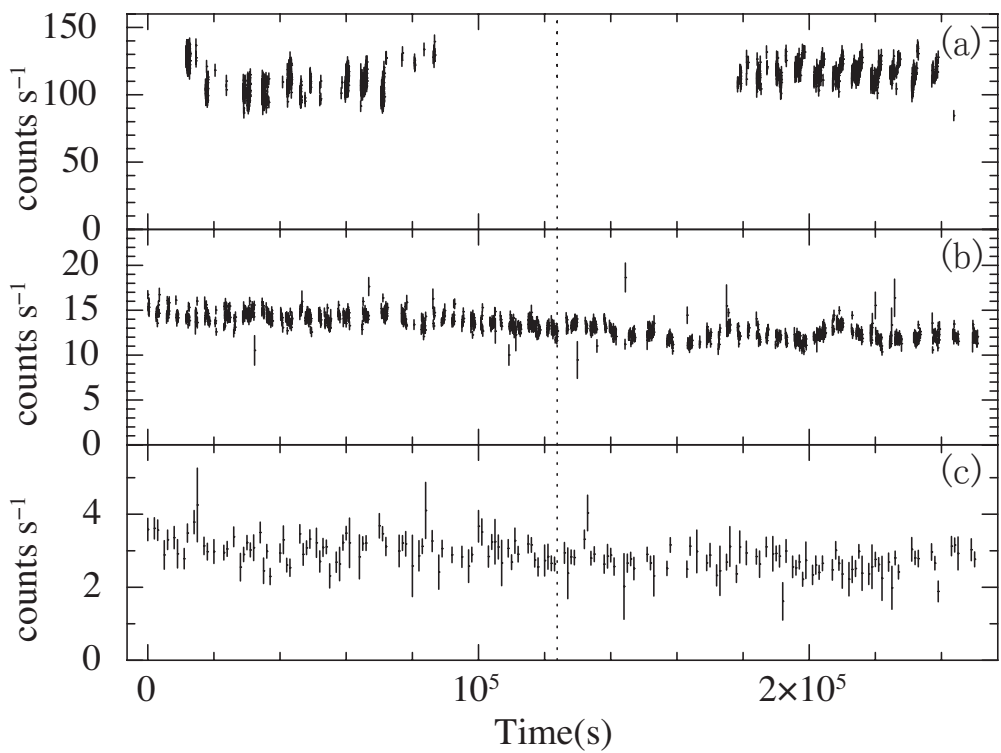}
   \caption{The Suzaku light curves of GX~$339-4$ during the observation.
   Count rate of 0.7--10~keV XIS0 (panel a), 10--60~keV PIN (panel b), and 50--200~keV GSO (panel c) were plotted 
   against time since 2007 February 12th 05:33:31.
   The XIS0, PIN, and GSO light curves are binned in 6~sec, 100~sec, and 1000~sec, respectively.
   The XIS0 events within the central region of $3^\prime$ were included in panel (a).
   The PIN and GSO light curves were corrected for dead time, and the NXB was subtracted.
   A vertical dotted line divided the observation between the first half and the latter half.
}
  \label{fig:suzaku_lc}
\end{figure}

\begin{figure}
\epsscale{.80}
\plotone{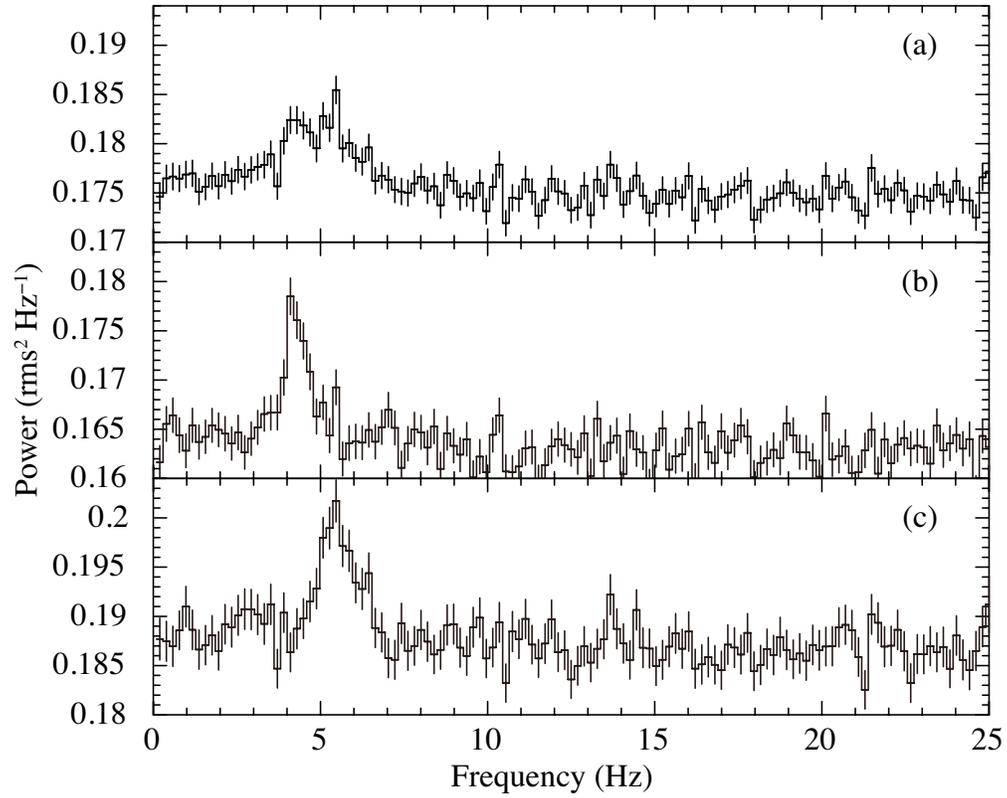}
  \caption{Power spectral density plots calculated based on NXB subtracted PIN lightcurve in the range of 10--60~keV, where white noise level was not subtracted.
Panel (a) shows the power spectral density based on the entire light curve, while panel (b) and (c) 
shows those of the first half  and the latter half, respectively. 
}
      \label{fig:powspec}
\end{figure}

Figure~\ref{fig:powspec}(a) shows the power spectral density from the entire PIN lightcurve calculated using 
{\sc xronos} (version5.22).
In this figure, a double QPO feature seen at frequencies of $\sim4$~Hz and $\sim5$--6~Hz.
During the observation, PIN and GSO count rates slightly decreased by
$\sim 25$\% and $\sim 18$\%, respectively, 
while that of XIS-0 slightly increased $\sim 9$\% (figure~\ref{fig:suzaku_lc}). 
Thus the double QPO feature is probably due to a
change of a single QPO frequency.

We reanalyzed the same data by dividing them into first half and latter half as indicated in 
figure~\ref{fig:suzaku_lc}.
The calculated power spectral densities were also shown in figure~\ref{fig:powspec}(b)(c)
showing a clear change in QPO frequency from 
$4.26\pm0.06$~Hz in the first half to 
$5.46\pm 0.06$~Hz, in the second half. 
Since these QPO frequencies are probably set by the size of the emission region, we look
for changes in the size of the disk and corona in their corresponding spectra. 
Specifically, a higher frequency QPO should indicate a smaller size region, and our
modeling with {\sc dkbbfth} means that we can track both the size of the disk
and the size of the corona via the parameters $r_{\rm in} ^*$ and $r_{\rm tran}$. We then 
compare the  observed changes with the predictions of 
two specific QPO models. 

\subsection{Time evolution of spectral parameters}

\begin{figure}
\epsscale{.80}
\plotone{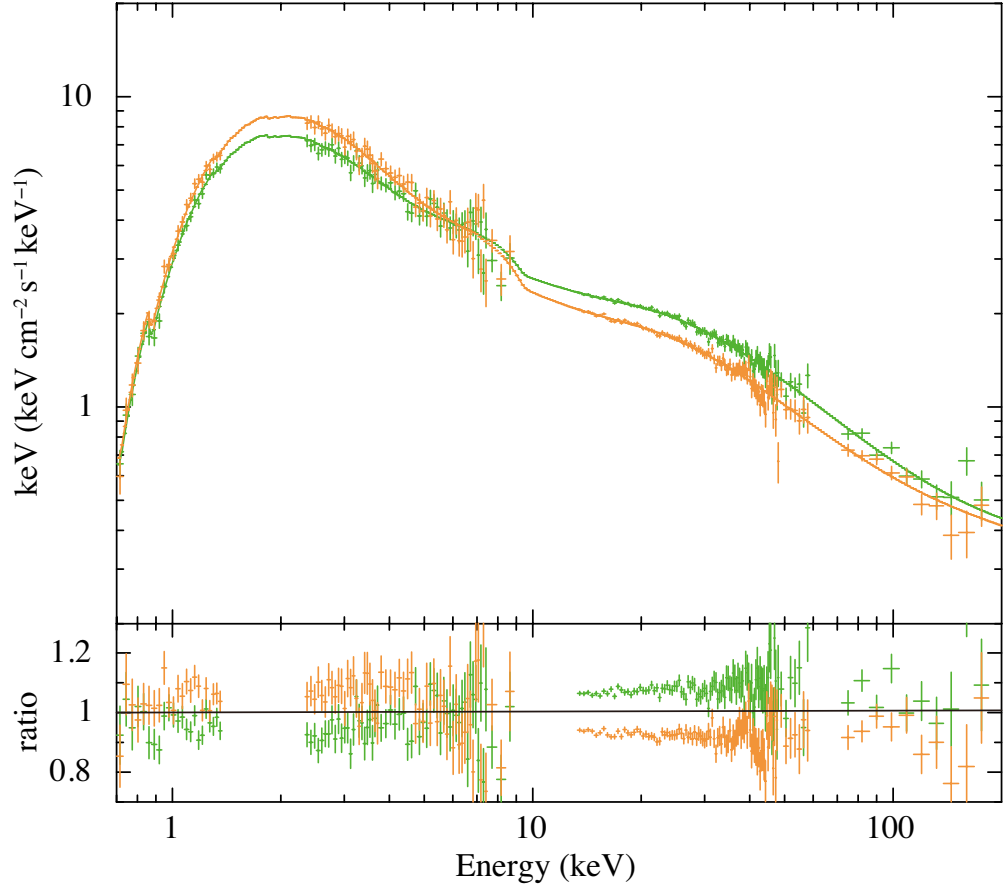}
  \caption{$\nu F_\nu$ spectra taken from the first half (green) and the latter half (orange) are plotted in the top panel based on the best fit model~4. The ratios to the 
   best fit model~4 for the summed spectrum are plotted in the same color in the bottom panel.}
  \label{fig:spec_each}
\end{figure}

To investigate evolution of the disk and corona in detail, 
we accumulated spectra from the two halves of the observation. 
For PIN and GSO data,  net exposures of 43.9~ks and 44.0~ks were acquired for the first half 
and the latter half, 
respectively. The NXB spectra were calculated for each periods and subtracted from the data. 
For XIS0 data, 1.6~ks and 1.3~ks exposures were acquired for the first half and the latter half,
respectively.
The two data sets have very similar
observed luminosities. The 
absorbed flux was almost same as 
$2.7\times 10^{-8}~{\rm erg~s^{-1}cm^{-2}}$ and 
$2.8\times 10^{-8}~{\rm erg~s^{-1}cm^{-2}}$, for the first half and the latter half, 
respectively, while their spectral shape was clearly different.
The top panel of figure~\ref{fig:spec_each} shows $\nu F_\nu$ spectra of the first half 
(green) and the latter half (orange), and
the bottom panel shows ratios of each spectrum to the best fit model~4 for 
the summed spectrum. There is a clear 
anti-correlation between the soft band data and the hard band data.  
The XIS0 data in 1--4~keV increased, while 
the PIN data decreased, so the spectrum pivots around 10~keV, becoming significantly
softer during the Suzaku observation. 

A closer inspection of figure~\ref{fig:spec_each} reveals that the 
data below 1~keV do not take part in this spectral pivoting, but remain relatively stable,
despite the 10--20~\% flux change in the range of 2--4~keV.
These photons below $\sim 1$~keV are from the outer part of the disk, 
where the  local temperature is lower than $\sim0.3$~keV. Since this emission does
not change it seems most likely that the mass accretion rate through the disk is
not changing (as is also implied by the constant luminosity), 
and it is only the geometry of the inner disk/corona which gives the spectral change above 1~keV.

In order to discuss change of the spectral parameters, 
we fit the two data sets with the independent corona (model~3) and the 
coupled corona (model~4). We  
fixed the values of $N_{\rm H}$, iron abundance, and the gaussian central energy
at those of the best fit values obtained by  the summed spectral fit (table~\ref{tab:bestfit_sum}). 
Furthermore, in the case of model~4, to avoid a strong coupling between 
$\tau$
and $kT_{\rm e}$, we fix $kT_{\rm e}$ to the value seen in the average spectrum. 

The results are summarized in table~\ref{tab:bestfit_12}, and again the 
energetically coupled corona model 
{\sc dkbbfth} gives the best fit. 
Based on the best fit model~4, 
the photon index of thermal corona $\Gamma_{\rm th}$ increases 
from $2.59\pm0.03$ to $2.74\pm0.04$. Since we keep $kT_{\rm e}$ constant between the two
datasets, this means that the 
optical depth of the thermal corona $\tau$ 
slightly decreased from $0.61\pm 0.02$ to $0.55^{+0.02}_{-0.01}$. Hence the second 
half of the data have a lower $y$ parameter and softer Compton tail, as seen in 
Figure~\ref{fig:spec_each}(a). This means that $f_{\rm th}$ also decreases, from 0.30 to 0.25.
The unComptonized disk emission is slightly more 
prominent in the second spectrum, so the transition radius decreases
slightly from $r_{\rm tran}=(2.6^{+0.4}_{-0.2}) r_{\rm in} ^*$ to $(2.3^{+0.4}_{-0.2}) r_{\rm in} ^*$. 

The intrinsic disk temperature $kT_{\rm in} ^{\rm int}$ marginally increases from 
$0.66\pm0.03$~keV to $0.69^{+0.02}_{-0.03}$~keV, giving an 
apparent inner radius of the underlying disk,  $r_{\rm in} ^\ast $
which decreases by $\sim 5\%$ from 
$(95^{+7}_{-6})\cdot \zeta_{50}d_8$~km to $(90^{+6}_{-4})\cdot \zeta_{50}d_8$~km. Thus the
outer radius of the corona, $r_{\rm tran}$ decreases by  
$\sim 16\%$ from $250\cdot \zeta_{50}d_8$~km to $210\cdot \zeta_{50}d_8$~km.
Model~3 gave a similar result that the disk inner radius 
decreased $\sim 5\%$ from the first half to the second half data.

\subsection{Discussion on the relation of QPO and the thermal corona}

Lense-Thirring precession  has been proposed as the origin of the 
low-frequency QPOs (e.g. \citealt{stella98}). This can explain both the
frequency and the spectrum of the QPO if the precession is of a hot inner
Comptonizing flow rather than the thin disk
\citep{ingram09}. In this model 
the QPO frequency $f_{\rm QPO}$ is   $f_{\rm QPO}\propto r^{-2.1}$.
Thus, a $28\pm3$\% increase in QPO frequency 
means that the  size of the precessing region should decrease by $\sim11\pm1$\%.
Since the precession is a vertical mode, the only part of the corona which can 
precess in this way is the corona interior to $r_{\rm in} ^*$ (see the schematic diagram in 
figure~\ref{fig:model}). The fits imply that $r_{\rm in} ^*$ decreases by $\sim 5\pm 8$\%,
consistent with the Lense-Thirring QPO model.
%

Another model for the low frequency QPO is that it is a mode
of the hot inner flow.  Global 3D magnetohydrodynamic (MHD) simulations 
suggest that the magnetorotational instability (MRI) may lead to
the quasi-periodic deformation of a hot inner torus 
from a circle to a crescent \citep{machida08}. 
The frequency in their simulation can be roughly related to the size of the 
inner torus as
$f_{\rm QPO}\propto r^{-3/2}$.
So the observed change of QPO frequency requires that the radius  change by $\sim15$\%.
This is slightly larger than the observed decrease in $r_{\rm in} ^*$.
However, their simulation was for an unconstrained hot flow, i.e. there was
no cool disk present. It is not clear how this will 
change the formation and deformation of such a hot inner torus. 
So the model is probably consistent within the systematic uncertainty.

\section{Summary and conclusions}

We have analyzed the Suzaku spectrum of GX~$339-4$ in the 
strongly Comptonized VHS/HIMS. We used a series of models to fit the data, 
starting with the commonly used
'diskbb plus power-law' model (model~1). We then use progressively better
models of the Comptonization, first assuming that it is non-thermal
(model 2) and then allowing it to be both thermal and non-thermal (model 3)
as required from previous studies (e.g \citealt{zdz01, gd03}). 
The strong Comptonization means that many seed photons from the disk 
are removed by Compton scattering so the disk luminosity
is larger than the directly observed luminosity. This effect increases the 
apparent radius of the disk, though using proper Comptonization models
rather than a power law already increases the disk luminosity and gives an
apparent radius which is larger than that seen in the HSS. 
Our final model couples the disk and corona together (model~4). 

Both these models except for model~4 assume that the disk and 
corona are independent components, 
yet both ultimately derive from the gravitational energy. Hence our
final model is a more physical one where an inner corona reduces the 
power available for the inner disk. This more physical model gives the 
best fit to the data, and again requires that the apparent radius of the 
disk is larger than that seen in the HSS. This increase in apparent size of the disk
can be seen directly by comparison with an XMM-Newton/RXTE HSS spectrum 
at almost the same luminosity 
taken 3 days after the Suzaku data. This clearly shows that the disk temperature
is lower in the VHS/HIMS than in the HSS. 
The only way that this can be 
consistent with the same true inner disk radius is if the color temperature 
correction and/or stress at the inner boundary decreases in the VHS/HIMS. 
However, it is much more likely that these actually increase, as there is 
strong irradiation in the VHS/HIMS which should increase the color temperature
correction, and it is not possible to decrease the stress on the inner boundary from 
the zero stress condition used for the HSS. Thus the data show that the
VHS/HIMS is associated with a truncated cool disk geometry, where the 
disk inner radius is 1.3--2.2 times larger than that of the constant radius 
seen in the HSS which is identified with the ISCO. 

Within the Suzaku observation there is a small change in flux and spectral
shape, and a corresponding change in the fast variability properties as 
determined from the power spectrum. The low frequency QPO increases 
from $4.26\pm0.06$~Hz to $5.46\pm0.06$~Hz as the spectrum softens, with a derived decrease in 
apparent disk inner radius of  $r_{\rm in} ^*=(95^{+7}_{-6})\cdot \zeta_{50}d_8$~km to 
$(90^{+6}_{-4})\cdot \zeta_{50}d_8$~km. While
this is not significant, this decrease in radius is consistent with the 
increase in QPO frequency if the QPO is formed either from Lens-Thirring
precession of the hot flow interior to $r_{\rm in}$ 
\citep{ingram09} or from a magneto-hydrodynamic mode of this flow
\citep{machida08}. 

A slightly truncated disk in the VHS/HIMS clearly makes a smooth connection to 
a larger radius disk truncation as required for LHS models which use the 
alternative hot flow solutions \citep{esin97}, making a coherent 
picture of the outbursts of BHB in a model where the 
spectrum changes from LHS-VHS/HIMS-HSS driven by a decreasing inner disk radius
until it reaches the ISCO. 

\bigskip
We would like to thank the referee Dr. Andrzej Zdziarski for his helpful comments.
We are grateful to all the Suzaku team members.
We also thank Mami Machida and Ryoji Matsumoto for helpful discussion on QPO and MHD.
The present work is supported in part 
by Grant-in-Aid No.19740113 from Ministry of Education, Culture, Sports, Science and Technology of Japan, 
and by Grant-in-Aid for Project Research in Shibaura Institute of Technology.

%

\appendix

\clearpage
\begin{deluxetable}{ll}
\tabletypesize{\scriptsize}
\tablecaption{Description of models in {\sc xspec}  \label{tab:models}}
\tablewidth{0pt}
\tablehead{
\colhead{model} & \colhead{descriptions in {\sc xspec} } 
}
\startdata
    model 1 & {\sc wabs*(diskbb+rdblur*ireflect*pl + gauss)}\\
    model 2\tablenotemark{a} & {\sc wabs*(simpl(d)*diskbb+rdblur*ireflect* simpl(c)*diskbb+gauss)}\\
    model 3\tablenotemark{a} & {\sc wabs*(simpl(d)*diskbb+rdblur*ireflect*(simpl(c)*diskbb+nthcomp)+gauss)}  \\
    model 4\tablenotemark{a,b} & {\sc wabs*(simpl(d)*dkbbfth(d)+rdblur*ireflect*(simpl(c)*dkbfth(d)+dkbbfth(c))+gauss)} \\    	
\enddata
       \tablecomments{To use convolution models, {\sc simpl}, {\sc ireflect}, and {\sc rdblur}, energies are extended in the range of  0.1--1000~keV.}    
      \tablenotetext{a}{{\sc simpl(d)} and {\sc simpl(c)} represent the direct and comptonized 
       component of the {\sc simpl} model, respectively.}
      \tablenotetext{b}{{\sc dkbbfth(d)} and {\sc dkbbfth(c)}  represent the direct and comptonized component of the {\sc dkbbfth} model, respectively.}
\end{deluxetable}

\begin{deluxetable}{llcccc}
\tabletypesize{\scriptsize}
  \tablecaption{The best fit parameters for the summed spectrum  \label{tab:bestfit_sum}}
\tablewidth{0pt}
\tablehead{
\colhead{Component} & \colhead{Parameter} & \colhead{model~1} & \colhead{model~2} & \colhead{model~3} & \colhead{model~4} 
}
\startdata
{\sc wabs}& $N_{\rm H}(10^{21}~{\rm cm^{-2}})$&
$6.7\pm0.2$ &$4.8^{+0.1}_{-0.2}$ &$4.9\pm0.2$ & $4.9^{+0.1}_{-0.2}$ \\ \tableline
{\sc diskbb}& $kT_{\rm in}$~(keV) &
$0.65\pm0.02$ &$0.55^{+0.02}_{-0.01}$ &$0.56\pm0.02$ & ---\\
&$r_{\rm in}$~($\zeta_{50}d_8$~km)&
$56^{+6}_{-5}$ & $122^{+7}_{-8}$ &$94^{+6}_{-5}$ &--- \\ \tableline
{\sc nthcomp} or {\sc dkbbfth} &$kT_{\rm in} ^{\rm int}$~(keV)&
  --- & --- &--- & $0.67^{+0.04}_{-0.08}$  \\
 &$\Gamma_{\rm th}$&  --- & --- &
 $2.72\pm0.04$ &$2.67^{+0.02}_{-0.03}$ \\
~~ &$kT_{\rm e}$ (keV)& --- & ---&
 $72^{+>230}_{-40}$ & $41\pm6$\\ 
&norm& --- & ---&$3.7\pm0.3$  & $1.2\pm0.3$\\
&$r_{\rm tran}$~($r_{\rm in}$)& --- & ---& --- & $2.5^{+0.4}_{-0.2}$\\ 
\cline{2-6}
&derived $\tau$ \tablenotemark{a} &---&---&$0.4\pm0.3$& $0.58^{+0.08}_{-0.07}$ \\
&derived $f_{\rm th}$&---&---&---& 0.27\\ \tableline
{\sc pl} or {\sc simpl}&$\Gamma_{\rm PL}$ &
$2.68^{+0.02}_{-0.01}$ &$2.68\pm0.01$ & (2.1) \tablenotemark{b}& (2.1) \tablenotemark{b}\\
&norm \tablenotemark{c} &$11.1\pm0.5$ & ---& --- &---\\ 
& $f_{\rm PL}$  &
---& $0.33\pm0.01$& $0.043^{+0.005}_{-0.009}$ & $0.032\pm0.002$\\ \tableline
{\sc ireflect}\tablenotemark{d}& $\Omega/2\pi$ &
$0.79\pm0.05$ & $0.82^{+0.03}_{-0.05}$& $0.83^{+0.07}_{-0.08}$ & $0.74^{+0.07}_{-0.07}$\\
&Fe abundance (solar)& 
$1.8\pm0.2$&$1.7\pm0.2$ & $1.8^{+0.4}_{-0.2}$ & $1.9^{+0.3}_{-0.2}$\\
\tableline
gaussian\tablenotemark{e}& E~(keV) & 
$6.94\pm0.34$& $6.96$\tablenotemark{f} &$6.94$\tablenotemark{f}  & $6.94$\tablenotemark{f}\\
 		&eqw~(eV) &
		$53^{+46}_{-45}$&$8^{+45}_{-8}$ & $5^{+46}_{-5}$ & $16^{+43}_{-16}$\\ \tableline
&$\chi^2/dof$ & 190.2/186 &198.9/186&192.0/184 & 178.7/184\\ \tableline 
\tableline
photon flux \tablenotemark{g}& $F_{\rm disk+th}^{\rm photon} ~({\rm photons~s^{-1}cm^{-2}})$&--- & ---& 39.7 & 43.5\\ 
inner radius\tablenotemark{h}			& $r_{\rm in} ^\ast$~($\zeta_{50}d_8$~km) &---&---&$117\pm5$&$93^{+19}_{-8}$\\ 
\enddata
\tablecomments{
    Errors represents 90\% confidence limit of statistical errors. Energy band was extended as 0.1--1000~keV.}
        \tablenotetext{a}{The optical depth calculated vie equation~(1).}
       \tablenotetext{b}{Fixed in the model fitting.}
        \tablenotetext{c}{${\rm photons~keV^{-1}cm^{-2} s^{-1}}$ at 1~keV}
        \tablenotetext{d}{The solar abundances are assumed for heavy elements except for iron.
       The temperature of the reflector and ionization parameter are fixed at $10^6$~K 
       and $300~{\rm erg~cm~s^{-1}}$, respectively.
       The reflected components are relativistically blurred by {\sc rdblur} with fixed $R_{\rm in} ^{\rm rdblur}=10~R_G$, $R_{\rm out} ^{\rm rdblur}=10^5~R_G$, $\beta=-3$, and $i=50^\circ$.}
        \tablenotetext{e}{Gaussian sigma is fixed at 0.2~keV.}
        \tablenotetext{f}{Value of the gaussian central energy $E$ is limited to be 6.2--7.8. Its 90\% confidence exceeds this range with model~2,3, and 4. }
       \tablenotetext{g}{ Unabsorbed photon flux of the disk and thermal Compton components in the range of 0.01--100~keV, where the reflected emission is excluded. The photon flux of {\sc diskbb} plus {\sc thcomp} and 
       that of {\sc dkbbfth} are shown for model~3 and model~4, respectively. }
        \tablenotetext{h}{The apparent disk inner radius estimated via equation~(A.1) in \cite{kubota04} for the slab geometry.}
\end{deluxetable}

\begin{deluxetable}{llcccc}
\tabletypesize{\scriptsize}
\tablecaption{The best fit parameters of model 3 and 4 for the spectra in the first half and the latter half\label{tab:bestfit_12}}
\tablewidth{0pt}
\tablehead{
\colhead{Component} & \colhead{Parameter} &\multicolumn{2}{c}{model3} &\multicolumn{2}{c}{model4} \\
\colhead{} & \colhead{} & \colhead{first half} & \colhead{latter half}& \colhead{first half} & \colhead{latter half}
}
\startdata
{\sc diskbb}& $kT_{\rm in}$~(keV) &$0.54\pm0.01$& $0.58^{+0.02}_{-0.01}$ & --- & --- \\
&$r_{\rm in}$~($\zeta_{50}d_8$~km)&$99^{+3}_{-4}$&$91\pm3$& --- & --- \\ 
\hline
{\sc nthcomp} or {\sc dkbbfth}&$kT_{\rm in} ^{\rm int}$~(keV) & --- & --- & $0.66\pm0.03$&$0.69^{+0.02}_{-0.03}$ \\
&$\Gamma_{\rm th}$   &$2.63^{+0.03}_{-0.04}$&$2.80^{+0.04}_{-0.05}$&$2.59\pm0.03$&$2.74\pm0.04$\\
&$kT_{\rm e}$~(keV) &$38^{+41}_{-12}$&$>41$&(41)\tablenotemark{a}&(41)\tablenotemark{a}\\
 &norm  &$3.5\pm0.3$& $3.7^{+0.3}_{-0.4}$&$1.3^{+0.3}_{-0.2}$&$1.2^{+0.1}_{-0.2}$\\
 &$r_{\rm tran}$~($r_{\rm in}$)& --- & --- &$2.6^{+0.4}_{-0.2}$&$2.3^{+0.4}_{-0.2}$\\
 \cline{2-6}
 &derived $\tau$  &$ 0.6\pm0.3 $&$<0.55$  &$0.61\pm0.02$&$0.55^{+0.02}_{-0.01}$\\
  &derived $f_{\rm th}$  & --- & --- &$0.30 $&$0.25$\\
\tableline
 {\sc simpl}&$\Gamma_{\rm PL}$   & (2.1)& (2.1)& (2.1)& (2.1) \\
&$f_{\rm PL}$ & $0.052^{+0.005}_{-0.006}$& $0.035^{+0.005}_{-0.004}$&$0.035\pm0.002$&$0.028\pm0.002$\\
\tableline
{\sc ireflect}& $\Omega/2\pi$  &$0.76^{+0.09}_{-0.09}$&$0.88^{+0.11}_{-0.04}$&$0.66^{+0.06}_{-0.08}$&$0.82\pm0.10$\\
&Fe abundance (solar)&(1.8)\tablenotemark{a}&(1.8)\tablenotemark{a}&(1.9)\tablenotemark{a}&(1.9)\tablenotemark{a}\\
\tableline
gaussian & E~(keV) &  (6.94)\tablenotemark{a}& (6.94)\tablenotemark{a} &(6.94)\tablenotemark{a}& (6.94)\tablenotemark{a}\\
 		&eqw~(eV)  &$<49$& $<64$&$<58$&$5^{+66}_{-5}$\\
\tableline		
		&$\chi^2/dof$ &185.1/184&190.9/184 &180.5/185&184.6/185\\
		\tableline
		\tableline
photon flux &$F_{\rm disk+th} ^{\rm photon}~({\rm photons~s^{-1}~cm^{-2}})$  &38.7&41.1 &43.0&44.3\\
inner radius &$r_{\rm in}^\ast$~($\zeta_{50}d_8$~km) &$122^{+4}_{-5}$&$116^{+3}_{-4}$&$95^{+7}_{-6}$& $90^{+6}_{-4}$\\
\enddata
   \tablecomments{Same as table~\ref{tab:bestfit_sum} but for the separated spectra fit with model~3 and 4.}
       \tablenotetext{a} { Fixed at the best fit values for the summed spectrum in table~\ref{tab:bestfit_sum}.    }
       \end{deluxetable}

\clearpage

\end{document}